\documentclass[prx,twocolumn,nofootinbib,citeautoscript,longbibliography,notitlepage,superscriptaddress]{revtex4-1}
\usepackage{csquotes}
\usepackage{graphicx}
\usepackage{dcolumn}
\usepackage{bm}
\usepackage{color}
\usepackage{amsmath}
\usepackage{tabularx}
\usepackage{epstopdf}
\usepackage{booktabs}
\usepackage{nicematrix}
\usepackage{latexsym}
\usepackage{amssymb}
\usepackage{amsmath}
\usepackage{colortbl}
\usepackage{psfrag}
\usepackage{bbm}
\usepackage{bm}
\usepackage{titlesec}
\usepackage{dsfont}
\usepackage{feynmp}
\usepackage{slashed}
\usepackage{multirow}
\usepackage[tight]{subfigure}

\usepackage{comment}

\usepackage[papersize={8.5in,11in}]{geometry}
\usepackage{color}
\definecolor{darkblue}{rgb}{0.,0.,0.4}
\definecolor{darkred}{rgb}{0.5,0.,0.}
\definecolor{BlueViolet}{RGB}{138,43,226}
\definecolor{SkyBlue}{RGB}{30,144,255}
\definecolor{DarkGreen}{RGB}{0,100,0}
\usepackage[pdftex,colorlinks=true,linkcolor=darkblue,citecolor=blue,urlcolor=darkred]{hyperref}

\renewcommand{\epsilon}{\varepsilon}

\allowdisplaybreaks[3] 
\usepackage{CJKutf8}

\usepackage{amsmath}
\usepackage{feynmp}
\usepackage{amssymb}
\usepackage{epsfig}
\usepackage{graphicx}
\usepackage{subfigure}
\usepackage{mathrsfs}
\usepackage{cleveref}
\usepackage{cases}
\usepackage{bbm}
\usepackage{xcolor}
\usepackage{tabularx}
\usepackage{array}
\usepackage{booktabs}
\usepackage{multirow}
\usepackage{lipsum}
\usepackage{float}
\usepackage{lmodern}
\usepackage{textcomp}
\usepackage{makecell}
\usepackage{comment}
\usepackage{threeparttable}
\usepackage{orcidlink}  

\allowdisplaybreaks[4]

\begin{document}
\begin{CJK*}{UTF8}{gbsn}
\title{Quasinormal modes of coupled metric-dilaton perturbations in two-dimensional stringy black hole}

\author{Wen-Hao Bian (边文浩)\,\orcidlink{0009-0005-9980-3376}}
\email[]{whbian@smail.nju.edu.cn}
\affiliation{School of Physics, Nanjing University, Nanjing, Jiangsu 210093, China}

\author{Zhu-Fang Cui (崔著钫)\,\orcidlink{0000-0003-3890-0242}}
\email[Contact author: ]{phycui@nju.edu.cn}
\affiliation{School of Physics, Nanjing University, Nanjing, Jiangsu 210093, China}

\date{\today}

\begin{abstract}
We investigate the quasinormal modes (QNMs) associated with intrinsic metric–dilaton coupled perturbations of the Mandal–Sengupta–Wadia (MSW) black hole in two-dimensional string theory. By suitable field redefinitions, we recast the gravity–dilaton system in terms of the conformal factor and a redefined dilaton field, reducing linear perturbation equations to coupled Schr\"odinger-type eigenvalue equations in the tortoise coordinate and solving them accordingly. All modes satisfy $\mathrm{Im}(\omega)<0$, confirming the linear stability of the MSW black hole under intrinsic coupled perturbations. Unlike external scalar-field perturbations, which yield purely imaginary frequencies, the intrinsic perturbations generically exhibit nonvanishing real parts, corresponding to oscillatory modes of the gravity–dilaton sector. The real part of the frequency displays a nonmonotonic dependence on the overtone number, while increasing the central-charge parameter $\sqrt{k}$ systematically decreases the damping rate and prolongs the relaxation time. These results indicate that intrinsic perturbations probe internal dynamical degrees of freedom and reveal characteristic features of the relaxation dynamics of two-dimensional stringy black hole.
\end{abstract}


\maketitle


\section{Introduction}

Black hole, as one of the most profound theoretical predictions of general relativity, have long placed their dynamical response and stability at the forefront of gravitational physics research~\cite{Wald1984book,Frolov1998book,Vasy2020CDM,Dafermos2025GRG,Berti2009CQG,Regge1957PR,Kokkotas1999LRR,Vishveshwara1970PRD,Teukolsky1973AJ}. A perturbed black hole does not instantly settle into a static state but gradually relaxes through a series of damped oscillations characterized by complex, discrete frequencies; these oscillatory patterns are known as quasinormal modes (QNMs)~\cite{Berti2009CQG,Kokkotas1999LRR,Vishveshwara1970PRD,Gayen2023GRG,Konoplya2011RMP,Nollert1999CQG,Franchini2024arXiv,Cardoso2019LRR,Bolokhov2025arXiv,Chen2021EPJP,Chen2019EPJC,Bian2026arXiv}. The associated complex frequencies are determined solely by the black hole's macroscopic parameters—such as mass, angular momentum, and charge—and are independent of the specific form of the initial perturbation, leading to their characterization as the black hole's ``characteristic fingerprint"~\cite{Gayen2023GRG,Mandal1991MPRA,Lin2024PRD}. In recent years, with the development of gravitational wave detection, QNMs have become a crucial bridge connecting theoretical black hole physics and observational gravitational wave astronomy~\cite{Gayen2023GRG,Witten1991PRD}. At a more fundamental level, a growing body of research suggests that the QNM spectrum may encode information about the black hole's microscopic structure and could play a key role in quantum gravity theories~\cite{Nojiri2001IJMPA,GRUMILLER2002PR,Collas1977tk,Shen2026IJMPD,Konoplya2011RMP,Berti2025arXiv}.

Among the plethora of models, two-dimensional dilaton black hole have garnered significant attention due to their foundational role in the low-energy effective theories of string theory. In particular, the MSW black hole, proposed by Mandal, Sengupta, and Wadia, stands as an important exact solution within the context of two-dimensional string theory~\cite{Gayen2023GRG,Blazquez2017PRD,Shu2004PRD}. This model not only provides a controllable platform for investigating black hole thermodynamics and microstate counting but also frequently serves as a simplified framework for exploring quantum gravity effects and the holographic principle. Unlike four-dimensional Einstein gravity, pure two-dimensional Einstein gravity is a topological theory and possesses no local propagating gravitational degrees of freedom~\cite{JACKIW1985NPB,Teitelboim1983PLB,Boozer2008EJP,Jackiw1995arXiv}. However, within the string theory framework, the dilaton field is deeply coupled with the geometry, introducing a propagating scalar degree of freedom into the two-dimensional spacetime. This allows two-dimensional black hole to support nontrivial dynamical perturbations~\cite{Vieira2026Universe}.

Existing research has primarily focused on the response of the MSW black hole to perturbations by external test fields~\cite{Gayen2023GRG,Gursel2019MPLA}. For instance, Gayen and Koley systematically investigated the QNM spectra of external non-minimally coupled scalar and spinor fields in the background of a two-dimensional dilaton black hole. They found that scalar field perturbations yield purely imaginary frequencies, while spinor perturbations exhibit complex frequencies with nonzero real parts, whose magnitude is determined by the coupling constant and remains largely independent of the overtone number~\cite{Gayen2023GRG}. Furthermore, Gursel et al. studied the linear stability of the MSW black hole, demonstrating, by constructing Schr\"odinger-type wave equations, that the solution is stable under external perturbations~\cite{Gursel2019MPLA}. Sebastian and Kuriakose employed the WKB method to analyze Dirac QNMs in a (2+1)-dimensional MSW black hole, further confirming stability under perturbations~\cite{Sebastian2014MPLA}. In a broader context, Bl\'azquez-Salcedo et al. investigated coupled perturbations in Einstein-Gauss-Bonnet-dilaton black hole, discovering that dilaton coupling can significantly alter the QNM spectrum and potentially give rise to long-lived modes~\cite{Blazquez2017PRD,Blazquez2025PRD}. Similarly, in Garfinkle-Horowitz-Strominger black hole, dilaton coupling was also found to substantially influence the perturbation decay rate~\cite{Shu2004PRD}. These studies collectively underscore the significant role played by the dilaton field in black hole relaxation dynamics~\cite{Moura2023arXiv}.

However, the aforementioned studies largely treat the perturbations as external test fields, assuming a fixed black hole geometry and studying only the propagation and decay of the test field within that background. A more natural and physically profound question arises: what is the structure of the black hole's QNM spectrum if one directly perturbs the dynamical degrees of freedom of the black hole system itself---namely, perturbing the metric $g_{\mu\nu}$ and the dilaton field $\phi$ simultaneously? These intrinsic coupled perturbations, corresponding to QNMs that can be interpreted as the black hole's own ``vibrational modes" rather than responses to external probes, may offer a more direct reflection of the dynamical characteristics of the black hole's microscopic structure.

This paper aims to systematically investigate the QNM spectrum arising from coupled metric-dilaton intrinsic perturbations in the two-dimensional MSW black hole. Starting from the low-energy effective action of the MSW model, we employ appropriate field redefinitions and conformal transformations to reduce the perturbation equations to a set of coupled Schr\"{o}dinger-type matrix equations in the tortoise coordinate~\cite{Shen2026IJMPD}. Subsequently, imposing the boundary conditions of purely ingoing waves at the event horizon and purely outgoing waves at spatial infinity~\cite{Gursel2019MPLA}, we formulate the problem as a matrix barrier scattering eigenvalue problem, solved numerically to obtain the complex frequency spectrum~\cite{Ma2024PRD}.

Our computational results reveal several important dynamical features of intrinsic coupled perturbations. First, all modes satisfy $\mathrm{Im}(\omega)<0$, confirming the dynamical stability of the MSW black hole under intrinsic perturbations. Second, unlike external scalar field perturbations which yield purely imaginary frequencies~\cite{Gayen2023GRG}, intrinsic coupled perturbations generate complex frequencies with non-vanishing real parts, indicating that the coupling between the metric and dilaton can excite oscillatory dynamics. Furthermore, we find that the real part of the frequency exhibits a non-monotonic behavior as a function of the overtone number: it increases with the mode order for low overtones but gradually decreases for higher overtones. This phenomenon reflects a competition between coupled oscillations and horizon dissipation. Additionally, increasing the central charge parameter $\sqrt{k}$ lowers the effective potential barrier, thereby reducing the damping rate and prolonging the perturbation lifetime.

These results demonstrate that the black hole QNM spectrum is not only dependent on the black hole parameters but also highly sensitive to the physical nature of the perturbation. Different types of perturbations probe the black hole's dynamical structure from distinct perspectives: external scalar fields primarily reveal purely dissipative relaxation processes, while intrinsic coupled perturbations uncover the cooperative dynamics between the geometric and dilaton degrees of freedom. Therefore, investigating intrinsic coupled perturbations provides a new dynamical angle for understanding the microscopic structure of string theory black hole.

The rest of this paper is organized as follows. Sec.~\ref{secII} introduces the MSW black hole solution and its representation in conformal coordinates. Sec.~\ref{secIII} derives the coupled metric-dilaton perturbation equations and casts them into a matrix Schr\"odinger form. Sec.~\ref{secIV} analyzes the asymptotic properties of the effective potential and establishes the quasinormal mode boundary conditions. Sec.~\ref{secV} presents the numerical results and discusses the spectral structure. Sec.~\ref{secVI} discusses the physical implications of our results and their possible connections to black hole microstructure. Finally, Sec.~\ref{secVII} provides a summary.

\section{The MSW black hole and intrinsic perturbations}\label{secII}

Einstein's gravity is purely topological in two-dimensional spacetimes, possessing no physical propagating degrees of freedom~\cite{JACKIW1985NPB,Teitelboim1983PLB,Boozer2008EJP,Jackiw1995arXiv}. However, the presence of an additional structure such as the dilaton field introduces a propagating scalar degree of freedom, which can excite dynamical modes rather than gauge artifacts~\cite{Gayen2023GRG,Mandal1991MPRA,Witten1991PRD,Callan1992PRD,Russo2992PRD}. Therefore, we commence from the low-energy effective action of the dilaton black hole in string theory, the Mandal-Sengupta-Wadia (MSW) action~\cite{Gayen2023GRG,Mandal1991MPRA,Witten1991PRD,Callan1992PRD,Russo2992PRD},
\begin{equation}\label{eq:action}
S = \int d^{2}x\sqrt{-g} e^{-2\phi}\left[R + 4(\nabla \phi)^{2} + 4\Lambda^{2}\right],
\end{equation}
where $R$ is Ricci scalar and we adopt geometric units remain the gravitational constant and the speed of light satisfy \(G = c = 1\). Here, \(g\) is the determinant of the metric, \(\phi\) denotes the dilaton field, and \(\Lambda^{2}\) represents the cosmological constant. The variation of the metric determinant is given by \(\delta \sqrt{- g} = (\sqrt{- g} /2) g^{\mu \nu}\delta g_{\mu \nu}\). Applying the variational principles \(\delta S / \delta g_{\mu \nu} = 0\) and \(\delta S / \delta \phi = 0\) yield the equations of motion for the metric \(g_{\mu\nu}\) and the dilaton field \(\phi\)~\cite{GRUMILLER2002PR},
\begin{equation}\label{eq:eom}
\left\{ \begin{array}{l}
\nabla_{\mu}\nabla_{\nu}\phi -g_{\mu \nu}\left[\nabla^{2}\phi -(\nabla \phi)^{2} + \Lambda^{2}\right] = 0 \\[4pt]
R + 4\nabla^{2}\phi -4(\nabla \phi)^{2} + 4\Lambda^{2} = 0
\end{array} \right.
\end{equation}
The static black hole solution and the corresponding dilaton solution are given by~\cite{Gayen2023GRG}
\begin{equation}\label{eq:MSWsoln}
ds^{2} = \left(1 - \frac{M}{r}\right)dt^{2} - \frac{k\, dr^{2}}{4r^{2}\left(1 - \frac{M}{r}\right)},\quad e^{-2\phi} = \frac{r}{M},
\end{equation}
where \(M\) is the mass of the black hole, and \(\sqrt{k} = 1 / \Lambda\) is the central charge parameter~\cite{Gayen2023GRG}. This solution was first obtained by Mandal, Sengupta, and Wadia in the context of two-dimensional string theory~\cite{Mandal1991MPRA}, and was subsequently profoundly interpreted by Witten from the perspective of conformal field theory~\cite{Witten1991PRD}. The Ricci scalar for this spacetime is \(R = - 4M / (kr)\). Our region of interest is the exterior of the event horizon \(r > M\) and the time coordinate \(-\infty < t < \infty\). At the event horizon \(r = M\), the metric components exhibit the behavior \(g_{tt} \to 0\) and \(g_{rr} \to \infty\). To handle this, we introduce the generalized tortoise coordinate \(r_{*}\)~\cite{Gayen2023GRG},
\begin{equation}\label{eq:tortoise}
dr_{*} = \frac{\sqrt{k}}{2}\frac{dr}{r\left(1 - M/r\right)} \quad \Rightarrow \quad r_{*} = \frac{\sqrt{k}}{2}\ln \left(\frac{r - M}{\sqrt{k}}\right).
\end{equation}
Consequently, we have \(r = M + \sqrt{k}\, e^{2r_{*}/\sqrt{k}}\). We further define the conformal factor~\cite{Gayen2023GRG}
\begin{equation}\label{eq:conformal_factor}
\Omega^{2}(r_{*}) = 1 - \frac{M}{r(r_{*})} = \frac{\sqrt{k}\, e^{2r_{*}/\sqrt{k}}}{M + \sqrt{k}\, e^{2r_{*}/\sqrt{k}}}.
\end{equation}
Utilizing this conformal factor, the black hole solution Eq.~\eqref{eq:MSWsoln} can be recast in a conformally flat form,
\begin{equation}\label{eq:conf_flat}
ds^{2} = \Omega^{2}(r_{*})(dt^{2} - dr_{*}^{2}).
\end{equation}
The central idea of this work is to treat the black hole not merely as a fixed background, but as a complete dynamical system. Therefore, we consider small intrinsic perturbations of the metric and the dilaton field around the background solution~\cite{Gursel2019MPLA,Liu2025arXiv,Blazquez2017PRD},
\begin{equation}\label{eq:perturbations}
\begin{array}{c}
g_{\mu \nu} = \bar{g}_{\mu \nu} + \epsilon h_{\mu \nu},\\[4pt]
\phi = \bar{\phi} + \epsilon \varphi,
\end{array}
\end{equation}
where \(\bar{g}_{\mu \nu}\) and \(\bar{\phi}\) represent the background solutions from Eq.~\eqref{eq:MSWsoln}, \(\epsilon \ll 1\) is a small bookkeeping parameter, and \(h_{\mu \nu}\) and \(\varphi\) denote the perturbations of the metric and dilaton, respectively. Given that two-dimensional gravity lacks propagating degrees of freedom, the metric is entirely determined by a local conformal factor. This implies that any metric can be written in a conformally flat form~\cite{GRUMILLER2002PR, Collas1977tk}. Accordingly, we adopt the following parameterization to describe the perturbed metric~\cite{GRUMILLER2002PR}:
\begin{equation}\label{eq:pert_metric_param}
ds^{2} = e^{2\rho}(dt^{2} - dr_{*}^{2}) = e^{2\rho}\eta_{\mu \nu}dx^{\mu}dx^{\nu},
\end{equation}
where \(\eta_{\mu \nu} = \mathrm{diag}(1, -1)\) is the two-dimensional Minkowski metric. The background metric can correspondingly be written as \(\bar{g}_{\mu \nu} = e^{2\rho_{0}}\eta_{\mu \nu}\), with
\begin{equation}\label{eq:rho0}
\rho_{0} = \frac{1}{2}\ln \Omega^{2} = \frac{1}{2}\ln \left(\frac{\sqrt{k}\,e^{2r_{*}/\sqrt{k}}}{M + \sqrt{k}\,e^{2r_{*}/\sqrt{k}}}\right).
\end{equation}
At this point, perturbations of the metric are fully encapsulated by perturbations of the conformal factor \(\rho\). We redefine the perturbation variables as follows,
\begin{equation}\label{eq:new_vars}
\left\{ \begin{array}{l}
\rho = \rho_{0} + \epsilon \psi, \\[4pt]
\phi = \phi_{0} + \epsilon \chi,
\end{array} \right.
\end{equation}
where \(\psi\) and \(\chi\) represent the perturbations of the conformal factor \(\rho\) and the dilaton field \(\phi\), respectively. To derive the dynamics of these perturbation variables, we need to rewrite the action using the new variables. For a conformally flat metric \(g_{\mu \nu} = e^{2\rho}\eta_{\mu \nu}\), the metric determinant is \(\sqrt{- g} = e^{2\rho}\), and the Ricci scalar simplifies to~\cite{Wald1984book,Christensen1977PRD,Carroll2019book},
\begin{eqnarray}
R = -2e^{-2\rho}(\partial_{t}^{2}-\partial_{r_{*}}^{2})\rho. \label{eq:R_scalar}
\end{eqnarray}
For the kinetic term of the scalar field, we have,
\begin{equation}\label{eq:kinetic_term}
(\nabla \phi)^{2} = g^{\mu \nu}\partial_{\mu}\phi \partial_{\nu}\phi = e^{-2\rho}(\partial \phi)^{2}.
\end{equation}
Substituting Eqs.~\eqref{eq:R_scalar} and \eqref{eq:kinetic_term} into the original action Eq.~\eqref{eq:action}, we obtain,
\begin{align}
S = \int d^{2}x e^{-2\phi}\left[-2\partial^{2}\rho+4(\partial\phi)^{2}+4\Lambda^{2}e^{2\rho}\right]. \label{eq:action_rho_phi}
\end{align}
Performing integration by parts on the first term (ignoring boundary terms) yields
\begin{align}
\int d^{2}x e^{-2\phi}(-\partial^{2}\rho)
=
-2\int d^{2}x e^{-2\phi}\partial_{\mu}\phi \partial^{\mu}\rho. \label{eq:byparts}
\end{align}
Thus, the action expressed in terms of \(\rho\) and \(\phi\) becomes,
\begin{equation}\label{eq:action_rho_phi_final}
S = \int d^{2}x e^{-2\phi}\left[4\partial_{\mu}\phi \partial^{\mu}\rho +4(\partial \phi)^{2} + 4\Lambda^{2}e^{2\rho}\right].
\end{equation}
To simplify the exponential coupling, we introduce a new variable \(\Phi \equiv e^{-\phi}\) (i.e., \(\phi = - \ln \Phi\)), whose background solution is
\begin{equation}\label{eq:Phi0}
\Phi_{0} = e^{-\phi_{0}} = \sqrt{\frac{r}{M}} = \left(1 + \frac{\sqrt{k}}{M} e^{2r_{*}/\sqrt{k}}\right)^{1 / 2}.
\end{equation}
Using the transformation relations \(\partial_{\mu}\phi = - \partial_{\mu}\Phi /\Phi\), \(e^{-2\phi} = \Phi^{2}\), and \((\partial \phi)^{2} = (\partial \Phi)^{2} / \Phi^{2}\), the action Eq.~\eqref{eq:action_rho_phi_final} can be rewritten in terms of \(\rho\) and \(\Phi\),
\begin{align}
S = 4\int d^{2}x\left[-\Phi\,\partial_{\mu}\Phi\,\partial^{\mu}\rho+(\partial\Phi)^{2}+\Lambda^{2}\Phi^{2}e^{2\rho}\right]. \label{eq:action_rho_Phi}
\end{align}
With this, we have completed the redefinition of the dynamical variables, laying the foundation for systematically deriving the coupled perturbation equations in the subsequent section. This method of manipulation is consistent with the field reparameterization techniques discussed in the review of two-dimensional dilaton gravity by Grumiller et al.~\cite{GRUMILLER2002PR}.

\section{Derivation of the coupled perturbation equations}\label{secIII}

Through the transformations in the previous section, we have reformulated the original problem of metric and dilaton perturbations into a dynamical analysis of the variables \(\rho\) and \(\Phi\). We now formally introduce perturbations of these two fields
\begin{equation}\label{eq:pert_rho_Phi}
\left\{ \begin{array}{l}
\rho = \rho_{0} + \epsilon \psi, \\[4pt]
\Phi = \Phi_{0} + \epsilon \chi,
\end{array} \right.
\end{equation}
where \(\chi\) is the perturbation of \(\Phi\), related to the original dilaton perturbation \(\varphi\) by \(\chi = -\Phi_{0}\varphi\). Substituting the perturbation decomposition Eq.~\eqref{eq:pert_rho_Phi} into the action Eq.~\eqref{eq:action_rho_Phi} and expanding to order \(\epsilon^{2}\), we obtain
\begin{equation}\label{eq:action_expansion}
S = S^{(0)} + \epsilon S^{(1)} + \epsilon^{2}S^{(2)} + \mathcal{O}(\epsilon^{3}),
\end{equation}
where the zeroth-order term \(S^{(0)}\) is $S[\Phi_0,\rho_0]$.
This yields the background field equations, which we assume are satisfied by the background solution. The zeroth-order term is essentially a constant and does not participate in the perturbation dynamics. Consequently, \(S^{(1)}\) should vanish (on-shell),
\begin{align}
S^{(1)} = &-4\int \bigg\{\left[(\Phi_{0}\partial_{\mu}\chi +\chi \partial_{\mu}\Phi_{0})\partial^{\mu}\rho_{0} + \Phi_{0}\partial_{\mu}\Phi_{0}\partial^{\mu}\psi \right] \nonumber \\
&- 2 \partial_{\mu}\Phi_{0}\partial^{\mu}\chi -2\Lambda^{2}e^{2\rho_{0}}(\Phi_{0}\chi +\Phi_{0}^{2}\psi)\bigg\}. \label{eq:S1}
\end{align}
It can be verified, using the background field equations, that \(S^{(1)}\) vanishes identically for all \(\psi,\,\chi\). This provides a consistency check for the background solution. Specifically, since the first-order term \(S^{(1)}\) is a linear functional of the perturbation fields \(\psi, \chi\), the variations satisfy
\begin{equation}\label{eq:var_zero}
\frac{\delta S}{\delta\Phi}\bigg|_{0} = \frac{\delta S}{\delta\rho}\bigg|_{0} = 0.
\end{equation}
Therefore, we have
\begin{equation}\label{eq:S1_zero}
S^{(1)} = \int d^{2}x\left(\frac{\delta S}{\delta\rho}\bigg|_{0}\psi +\frac{\delta S}{\delta\Phi}\bigg|_{0}\chi\right) = 0.
\end{equation}
Thus, the first-order term is identically zero and contains no new information; it merely restates the background field equations. For the quadratic term, we obtain
\begin{eqnarray}
S^{(2)} 
&=& 
\int d^{2}x\,\mathcal{L}^{(2)},\nonumber\\
\mathcal{L}^{(2)} 
&=& 
4(\partial \chi)^{2} - 4\Phi_{0}\,\partial_{\mu}\chi \,\partial^{\mu}\psi \nonumber\\
&&-4\chi \,\partial_{\mu}\chi \,\partial^{\mu}\rho_{0} - 4\chi \,\partial_{\mu}\Phi_{0}\,\partial^{\mu}\psi \nonumber\\
&&+4\Lambda^{2}e^{2\rho_{0}}\left(\chi^{2} + 4\Phi_{0}\chi \psi +2\Phi_{0}^{2}\psi^{2}\right).
\end{eqnarray}
In summary, the quadratic action is what we truly need. We will now derive the equations of motion for \(\chi\) and \(\psi\) from this quadratic action. First, we derive the variation \(\delta S / \delta \psi = 0\), which yields the equation of motion for \(\psi\),
\begin{equation}\label{eq:eom_psi}
\Phi_{0}\,\partial^{2}\chi +\chi \,\partial^{2}\Phi_{0} + 2\partial_{\mu}\Phi_{0}\,\partial^{\mu}\chi +4\Lambda^{2}e^{2\rho_{0}}(\Phi_{0}\chi +\Phi_{0}^{2}\psi) = 0.
\end{equation}
Next, we derive \(\delta S / \delta \chi = 0\) to obtain the equation of motion for \(\chi\),
\begin{equation}\label{eq:eom_chi}
2\partial^{2}\chi -\Phi_0\partial^2\psi - \chi \partial^{2}\rho_0-2\Lambda^{2}e^{2\rho_{0}}\chi -4\Lambda^{2}e^{2\rho_{0}}\Phi_{0}\psi = 0.
\end{equation}
Since the background is static, we assume a harmonic time dependence for the perturbations~\cite{Gayen2023GRG},
\begin{equation}\label{eq:harmonic_time}
\psi(t,r_{*}) = \psi(r_{*})e^{-i\omega t},\quad \chi(t,r_{*}) = \chi(r_{*})e^{-i\omega t}.
\end{equation}
Substituting these into the equations of motion for \(\psi\) and \(\chi\), Eqs.~\eqref{eq:eom_psi}-\eqref{eq:eom_chi}, we obtain a set of coupled ordinary differential equations in the radial coordinate \(r_{*}\),
\begin{widetext}
\begin{align}\label{eq:coupled_ODE}
\left\{ \begin{array}{l}
\chi^{\prime \prime} + 2\frac{\Phi_{0}^{\prime}}{\Phi_{0}}\chi^{\prime} + \left[\omega^{2} + \frac{\Phi_{0}^{\prime\prime}}{\Phi_{0}} -4\Lambda^{2}e^{2\rho_{0}}\right]\chi -4\Lambda^{2}e^{2\rho_{0}}\Phi_{0}\psi = 0,\\[8pt]
2\chi^{\prime \prime} - \Phi_{0}\psi^{\prime \prime} + \left(2\omega^{2} - \rho_{0}^{\prime \prime} + 2\Lambda^{2}e^{2\rho_{0}}\right)\chi +\left(-\omega^{2}\Phi_{0} + 4\Lambda^{2}e^{2\rho_{0}}\Phi_{0}\right)\psi = 0.
\end{array} \right.
\end{align}
\end{widetext}
Here, a prime denotes differentiation with respect to the tortoise coordinate \(r_{*}\). It is worth noting that the coupled perturbation system corresponds to a non-normal Hamiltonian, which may lead to transient dynamical behavior~\cite{Besson2025FP}. The system of equations Eq.~\eqref{eq:coupled_ODE} is the starting point for our analysis of the quasi-normal modes of intrinsic perturbations in the MSW black hole. To transform it into a standard form more amenable to asymptotic analysis and numerical solution, we need to simplify it further. First, through a field redefinition \(\chi = \Phi_{0}^{-1}\tilde{\chi}\), we can eliminate the first-derivative term \(\chi^{\prime}\) in the first equation,
\begin{equation}\label{eq:chi_tilde_eq}
\tilde{\chi}^{\prime \prime} + \left(\omega^{2} - 4\Lambda^{2}e^{2\rho_{0}}\right)\tilde{\chi} -4\Lambda^{2}e^{2\rho_{0}}\Phi_{0}^{2}\psi = 0.
\end{equation}
Applying the same transformation to the second equation, while we obtain a system of second-order equations for \(\tilde{\chi}\) and \(\psi\), it may still contain first-derivative terms. In principle, this set of equations can be written in the following matrix form:
\begin{equation}\label{eq:matrix_form}
-\frac{d^{2}}{dr_{*}^{2}}\begin{pmatrix} \tilde{\chi} \\ \psi \end{pmatrix} + \mathbf{V}\begin{pmatrix} \tilde{\chi} \\ \psi \end{pmatrix} + \mathbf{W}\frac{d}{dr_{*}}\begin{pmatrix} \tilde{\chi} \\ \psi \end{pmatrix} = \omega^{2}\begin{pmatrix} \tilde{\chi} \\ \psi \end{pmatrix},
\end{equation}
where \(\mathbf{V}\) is the potential matrix, and \(\mathbf{W}\) is the first-derivative coupling matrix. For more complex modified gravity theories, it is sometimes difficult to reduce the perturbation equations to a matrix Schr\"odinger equation without first-derivative terms. To address such problems, some works have proposed alternative methods that directly analyze the asymptotic behavior of first-order differential systems~\cite{Langlois2021PRD}. In our case, it is possible to eliminate the \(\mathbf{W}\) matrix through an appropriate transformation. We seek an invertible transformation matrix \(\mathbf{P}\) such that the newly defined field \(\boldsymbol{\Psi} = \mathbf{P}\boldsymbol{\Phi}\), where \(\boldsymbol{\Psi} = (\tilde{\chi},\psi)^{\mathrm{T}}\), satisfies a standard matrix Schr\"odinger equation. By requiring the coefficient of the first-derivative term for the transformed field to vanish, we obtain a differential equation that \(\mathbf{P}\) must satisfy \(\mathbf{P}' = \mathbf{W}\mathbf{P} / 2\). Solving this equation yields:
\begin{equation}\label{eq:P_matrix}
\mathbf{P} = \begin{pmatrix} 1 & 0\\ \Phi_{0}^{-2} & 1 \end{pmatrix}.
\end{equation}
Under this transformation, we finally obtain a standard matrix Schr\"odinger equation without first-derivative terms:
\begin{equation}\label{eq:final_schrodinger}
-\frac{d^{2}}{dr_{*}^{2}}\boldsymbol{\Phi}(r_{*}) + V_{\mathrm{eff}}(r_{*})\boldsymbol{\Phi}(r_{*}) = \omega^{2}\boldsymbol{\Phi}(r_{*}).
\end{equation}
Here, \(V_{\mathrm{eff}} = \mathbf{P}^{-1}(-\mathbf{P}^{\prime \prime} + \mathbf{V}\mathbf{P} + \mathbf{W}\mathbf{P}^{\prime})\) is the final \(2\times 2\) effective potential matrix. The explicit expressions for its four matrix elements are
\begin{align}
V_{\mathrm{eff}}^{11,22} &= \frac{8\sqrt{k}\,e^{2r_{*}/\sqrt{k}}}{k\left(M + \sqrt{k}\,e^{2r_{*}/\sqrt{k}}\right)}, \label{eq:Veff1122} \\
V_{\mathrm{eff}}^{12} &= \frac{4V_{\mathrm{eff}}^{11}}{8 - k V_{\mathrm{eff}}^{11}}, \label{eq:Veff12} \\
V_{\mathrm{eff}}^{21} &= \frac{1}{128} V_{\mathrm{eff}}^{11}(8 - k V_{\mathrm{eff}}^{11})(32 - k V_{\mathrm{eff}}^{11}). \label{eq:Veff21}
\end{align}
At this stage, we have completely transformed the original physical problem into a matrix quantum mechanical barrier scattering problem. The QNM frequencies \(\omega\) of the system correspond to the eigenvalues of Eq.~\eqref{eq:final_schrodinger} under specific boundary conditions.

\section{Asymptotic analysis and boundary conditions}\label{secIV}

To impose the correct physical boundary conditions when solving the eigenvalue problem of Eq.~\eqref{eq:final_schrodinger}, we must analyze the asymptotic behavior of the effective potential at the boundaries \(r_{*} \to \pm \infty\). Since \(V_{\mathrm{eff}}\) is a coupling matrix, its specific form may change under different basis transformations. However, the true physical information, which is transformation-invariant, and its asymptotic behavior are determined by its eigenvalues \(\lambda_{\pm}(r_{*})\),
\begin{equation}\label{eq:eigenvalues}
\lambda_{\pm}(r_{*}) = \frac{1}{2}\left[\mathrm{Tr}V_{\mathrm{eff}}\pm \sqrt{(\mathrm{Tr}V_{\mathrm{eff}})^{2} - 4\det V_{\mathrm{eff}}}\right].
\end{equation}
By analyzing the behavior of the background functions at the boundaries, we can obtain the asymptotic values of the trace and determinant of \(V_{\mathrm{eff}}(r_{*})\). At asymptotic infinity \((r_{*} \to +\infty)\), we have \(\mathrm{Tr}V_{\mathrm{eff}} \to 16/k\) and \(\det V_{\mathrm{eff}} \to 16/k^{2}\). Consequently, the eigenvalues of the effective potential approach constants
\begin{equation}\label{eq:lambda_infinity}
\lambda_{\pm}^{\infty} = \frac{8}{k}(2\pm \sqrt{3}).
\end{equation}
These constants represent the squared effective masses of the two dynamical degrees of freedom in the asymptotic region. The presence of the dilaton field introduces a local scalar degree of freedom into the otherwise topological two-dimensional gravity and couples it to the metric. This coupling gives rise to non-zero effective masses in the asymptotic region, yielding qualitative results similar to those obtained from massive scalar field perturbations~\cite{Gayen2023GRG}. At infinity, the original perturbation fields \(\chi\) and \(\psi\) can be linearly combined to form \(\boldsymbol{\Phi}\), \(\lambda_{\pm}^{\infty}\) correspond to the asymptotic effective masses of the two modes, respectively. This feature is distinctly different from the massless gravitational waves of pure Einstein gravity and represents a unique dynamical signature of string-inspired gravity. Physically, we require purely outgoing waves at infinity (energy radiating outward), which corresponds to the boundary condition \(\boldsymbol{\Phi} \sim e^{+i\sqrt{\omega^{2} - \lambda_{\pm}^{\infty}}\,r_{*}}\)~\cite{Bekir2026arXiv}. It is worth noting that if the cosmological constant \(\Lambda \to 0\) (i.e., \(k \to \infty\)), then \(\lambda_{\pm}^{\infty} \to 0\), and the equation reduces to the wave equation for massless fields.

Near the horizon \((r_{*} \to -\infty)\), all background functions and potential matrix elements decay exponentially to zero. Therefore, the eigenvalues of the effective potential also approach zero, \(\lambda_{\pm} \to 0\). This is consistent with the standard result for black hole: the wave equation reduces to the free wave equation \(-\boldsymbol{\Phi}^{\prime \prime} = \omega^{2}\boldsymbol{\Phi}\). Physically, the horizon is a one-way membrane that allows only ingoing waves. Hence, we require the wave function to be purely ingoing at the horizon, corresponding to the boundary condition \(\boldsymbol{\Phi} \sim e^{-i\omega r_{*}}\).

Since the effective potential vanishes at the horizon, the two modes decouple there. Two independent solutions emerging from the horizon, when dynamically evolved to infinity, will generally mix both modes. The physically admissible general solution is a linear combination of these two solutions. The condition for the existence of QNMs is that there exists a non-zero set of coefficients such that the combined wave function contains only outgoing modes at infinity. Numerically, this condition can be implemented by constructing a matching matrix and requiring its singular values to vanish, which is a standard method for handling boundary conditions in QNMs calculations~\cite{Ma2024PRD,Berti2009CQG,Nollert1999CQG,Chandrasekhar1975PRSLAMPS}.

\begin{figure}
\centering
\subfigure[]
{\includegraphics[width=3in]{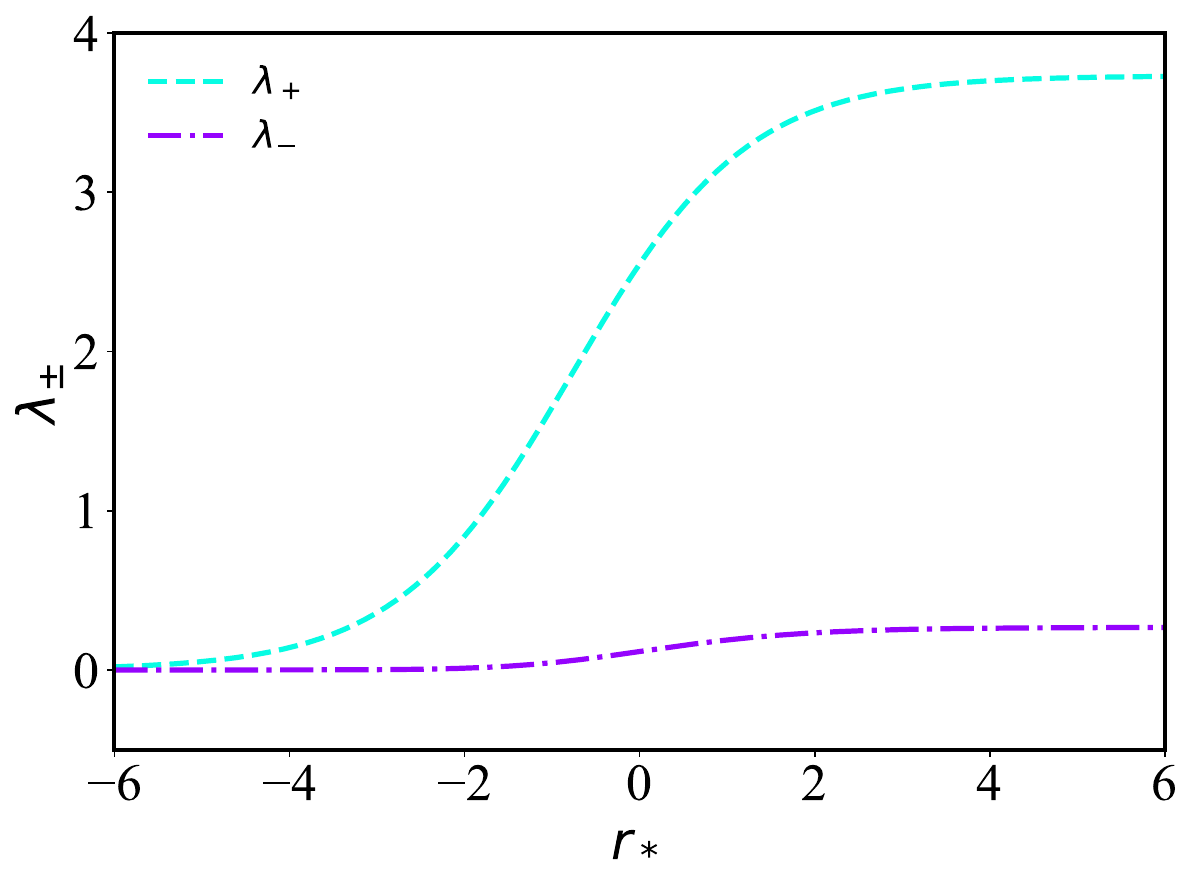}}
\subfigure[]{\includegraphics[width=3in]{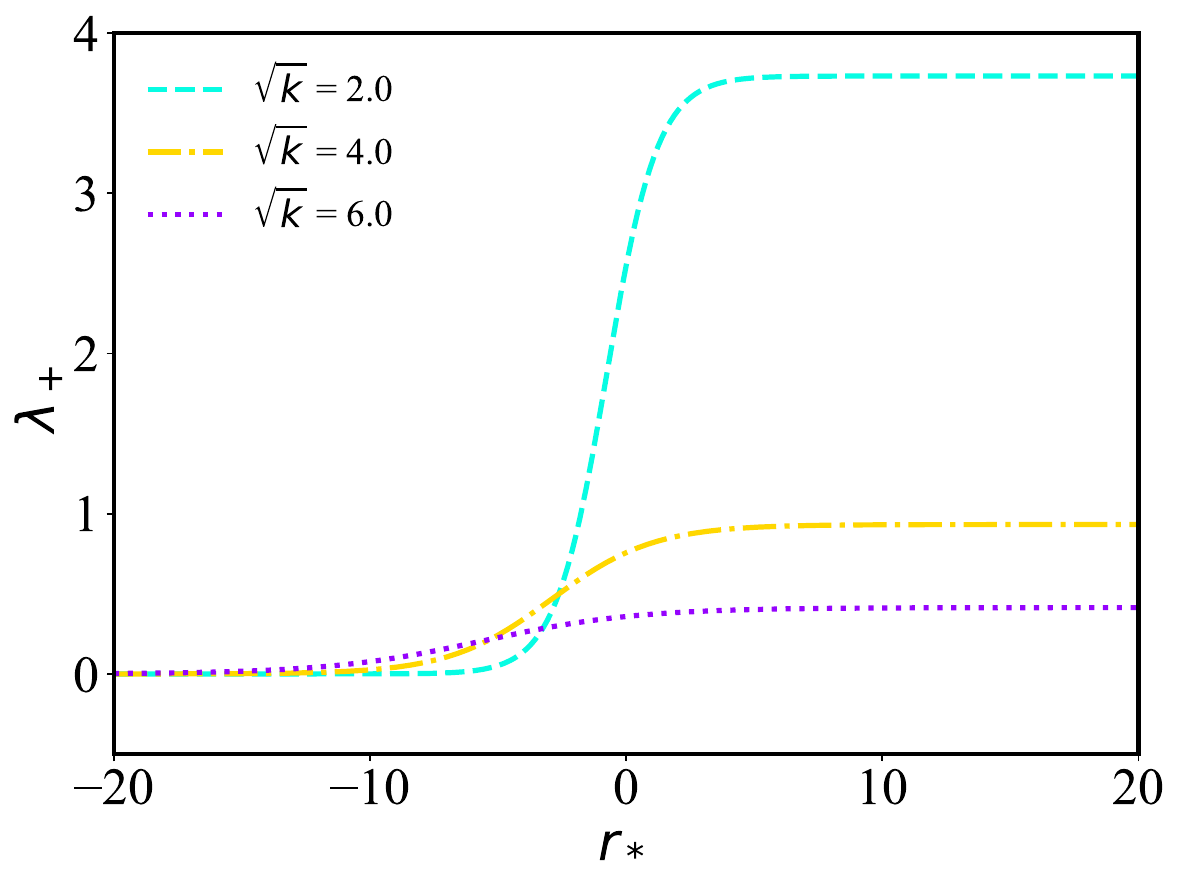}}
\caption{(Color online) (a) Evolution of the decoupled eigenvalues \(\lambda_{\pm}\) of the effective potential, Eq.~(\ref{eq:eigenvalues}), with \(M = 1.0\), \(\sqrt{k} = 2.0\); (b) Evolution of the branch \(\lambda_{+}\) for different values of \(\sqrt{k}\).}
\label{fig:effpot}
\end{figure}

To verify the boundary conditions of the effective potential and lay the foundation for subsequent numerical solutions, we analyze in detail the evolution of the effective potential eigenvalues \(\lambda_{\pm}\) as functions of the tortoise coordinate \(r_{*}\) from Eq.~(\ref{eq:eigenvalues}). Fig.~\ref{fig:effpot}(a) displays the global evolution of the two eigenvalues \(\lambda_{\pm}\) of the effective potential, which determines the dynamical behavior of the two propagation modes in different spatial regions. It is clearly observed that near the black hole horizon \(r_{*} \to -\infty\), both eigenvalues tend to zero, consistent with the asymptotic analysis above, indicating that the two modes propagate almost independently and freely near the horizon. As \(r_{*}\) increases, the two eigenvalues begin to separate and increase monotonically, eventually approaching different non-zero finite constants \(\lambda_{\pm}^{\infty} = 8/k(2\pm \sqrt{3})\) at infinity \(r_{*} \to +\infty\). This behavior implies that the two modes transition from nearly free propagation near the horizon to propagation dominated by effective masses in the asymptotic region, with their coupling determined by the spatial distribution of the off-diagonal elements.

Fig.~\ref{fig:effpot}(b) further explores the quantitative impact of the central charge parameter \(\sqrt{k}\) on the effective potential eigenvalue \(\lambda_{+}\). We fix the black hole mass \(M = 1.0\) and compare the cases \(\sqrt{k} = 2.0, 4.0, 6.0\). The results show that as \(\sqrt{k}/M\) increases (i.e., as the black hole mass decreases relatively), the saturation value of the effective potential at infinity decreases significantly. This finding carries profound physical implications: smaller black hole possesses lower effective potential barriers, meaning perturbation energy can more easily penetrate the barrier and propagate to infinity, rather than being localized and dissipated near the horizon, potentially prolonging the lifetime of perturbations~\cite{Gursel2019MPLA}. This provides crucial clues for our subsequent understanding of the variation of QNMs spectra with the central charge parameter as shown in Fig.~\ref{fig:effpot}(b). From the perspective of string theory, smaller black hole (with larger \(\sqrt{k}/M\)) are closer to the Planck scale, with a finite number of microstates and significant quantum effects. Therefore, the morphology of the effective potential and its parameter dependence may directly encode information about the black hole's microstate density or entropy corrections, opening up possible avenues for inferring microscopic structure from macroscopic dynamical responses.

In summary, we have imposed rigorous physical boundary conditions for solving the QNMs eigenvalue problem of Eq.~\eqref{eq:final_schrodinger}: purely ingoing waves at the horizon \((r_{*} \to -\infty)\) and purely outgoing waves at infinity \((r_{*} \to +\infty)\), with the specific form of the boundary conditions determined by the asymptotic behavior of the effective potential. This methodology for handling coupled perturbations follows the general theoretical framework for QNM studies~\cite{Nollert1999CQG}, laying a solid foundation for subsequent numerical solutions.

\section{Quasinormal modes and wave functions}\label{secV}

We have transformed the equations of motion for the intrinsic perturbations of the metric and dilaton fields into the standard Schr\"odinger equation form of Eq.~\eqref{eq:final_schrodinger}. Next, we proceed to numerically solve for its eigenvalues \(\omega^{2}\) to investigate the complex frequency spectrum of the QNMs and explore the evolutionary behavior of the 2D MSW dilaton black hole under intrinsic perturbations.

\begin{figure}
\centering
\subfigure[]
{\includegraphics[width=3in]{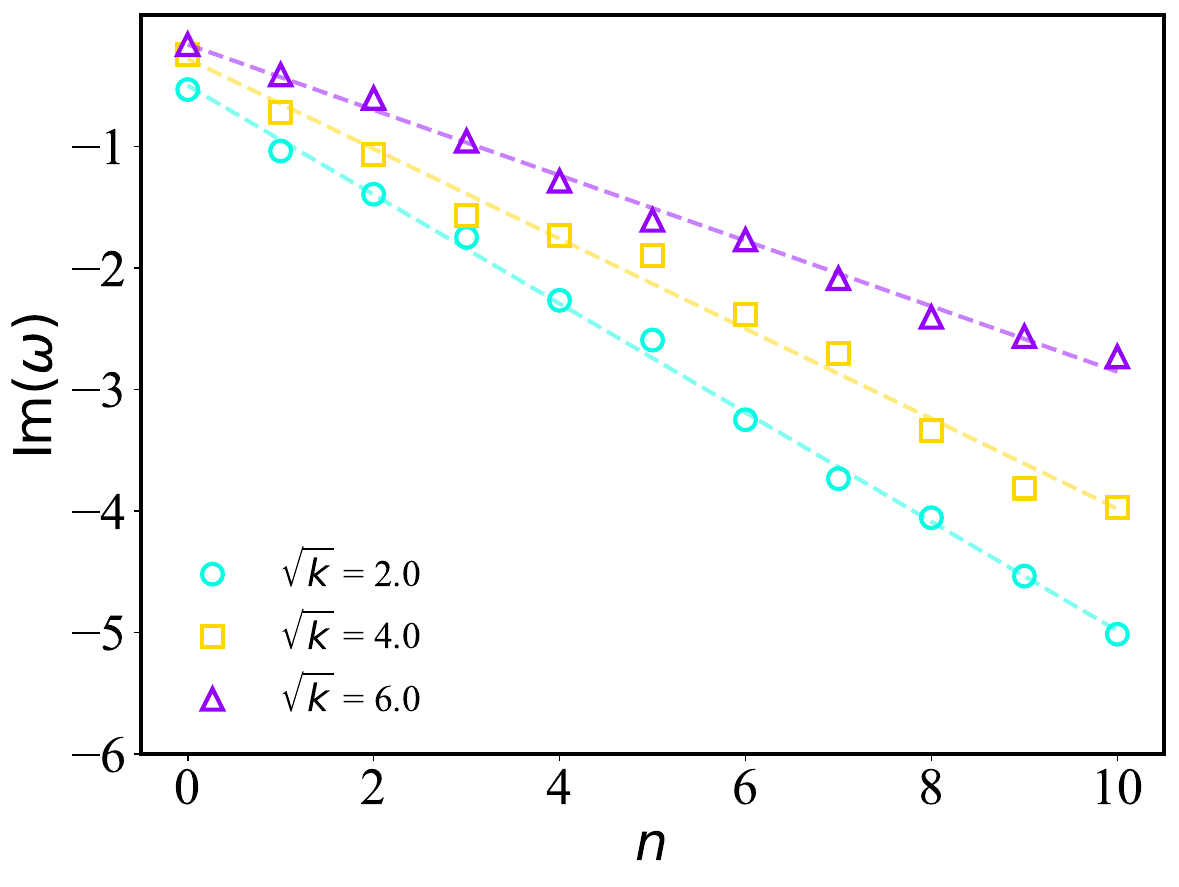}}
\subfigure[]
{\includegraphics[width=3in]{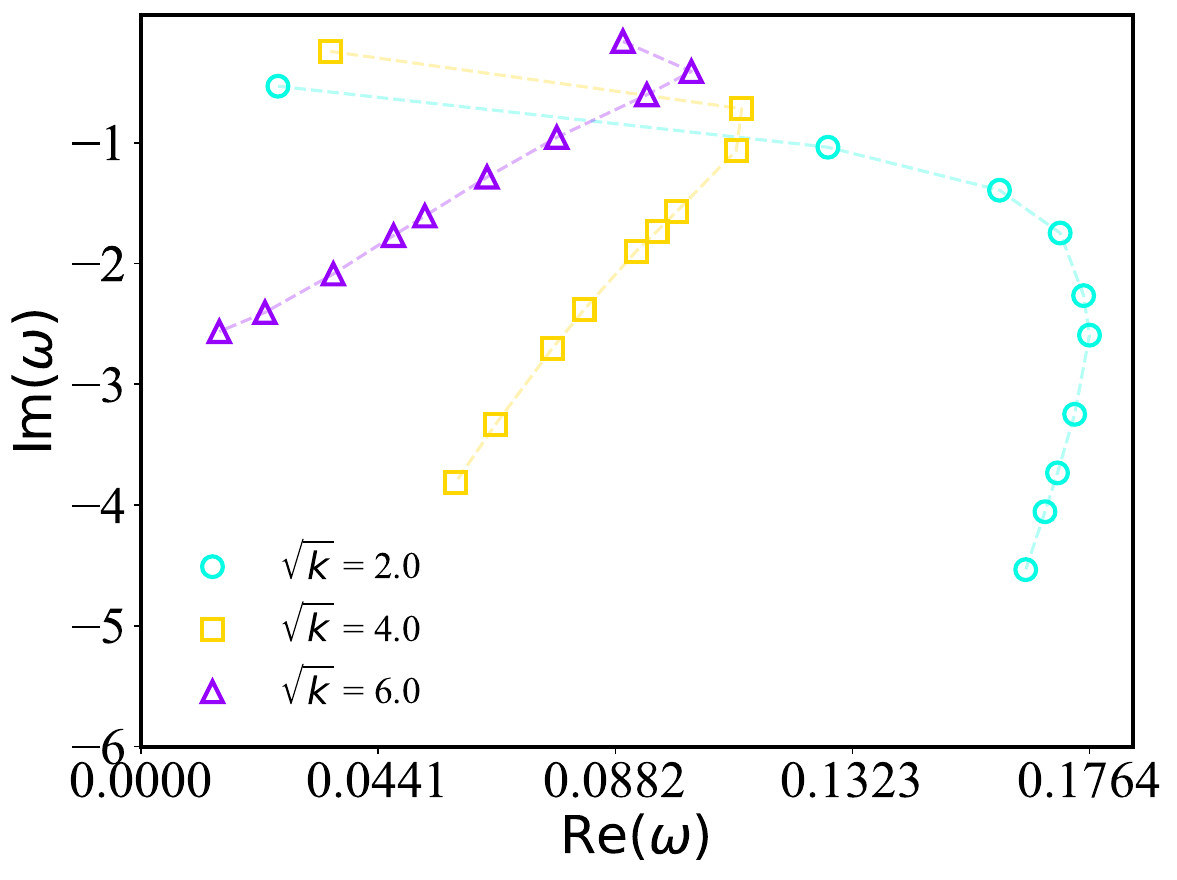}}
\caption{(Color online) Evolution of QNMs for parameters \(M = 1.0\) and several representative values of \(\sqrt{k}\): (a) Imaginary part \(\operatorname{Im}(\omega)\) vs. \(n\) and (b) Complex plane \(\operatorname{Im}(\omega)\) vs. \(\operatorname{Re}(\omega)\).}
\label{fig:qnm_evolution}
\end{figure}

\begin{table*}
\centering
\caption{Numerical values of black hole quasinormal modes (comparison of three sets of parameters), where decay time \(\tau = 1 / |\mathrm{Im}(\omega)|\), period \(T = 2\pi / \mathrm{Re}(\omega)\).}
\begin{tabular}{r||rrrr||rrrr||rrrr}
\hline
\hline
\multicolumn{4}{c}{$\sqrt{k}=2.0$} & \multicolumn{4}{c}{$\sqrt{k}=4.0$} & \multicolumn{4}{c}{$\sqrt{k}=6.0$} \\
\cmidrule(lr){2-5} \cmidrule(lr){6-9} \cmidrule(lr){10-13}
\hline
$n$& $\mathrm{Re}(\omega)$ & $\mathrm{Im}(\omega)$ & $T$ & $\tau$
& $\mathrm{Re}(\omega)$ & $\mathrm{Im}(\omega)$ & $T$ & $\tau$
& $\mathrm{Re}(\omega)$ & $\mathrm{Im}(\omega)$ & $T$ & $\tau$ \\
\midrule
0 & 0.024237 & -0.530719 & 11.8267 & 1.8842 & 0.034648 & -0.242975 & 25.6004 & 4.1156 & 0.089562 & -0.160408 & 34.2004 & 6.2341 \\
1 & 0.127177 & -1.036107 & 6.0188 & 0.9652 & 0.111659 & -0.717670 & 8.6509 & 1.3934 & 0.102319 & -0.409341 & 14.8914 & 2.4429 \\
2 & 0.159195 & -1.391989 & 4.4846 & 0.7184 & 0.113539 & -0.888537 & 7.0144 & 1.1254 & 0.094028 & -0.602738 & 10.2998 & 1.6591 \\
3 & 0.170486 & -1.746029 & 3.5815 & 0.5727 & 0.098980 & -1.571960 & 3.9891 & 0.6361 & 0.085775 & -0.780321 & 8.0038 & 1.2815 \\
4 & 0.174795 & -2.263725 & 2.7674 & 0.4417 & 0.095872 & -1.732087 & 3.6220 & 0.5773 & 0.064693 & -1.281562 & 4.8965 & 0.7803 \\
5 & 0.175142 & -2.595151 & 2.4156 & 0.3853 & 0.092000 & -1.897460 & 3.3075 & 0.5270 & 0.052577 & -1.607246 & 3.9072 & 0.6222 \\
6 & 0.174727 & -2.918371 & 2.1491 & 0.3427 & 0.082330 & -2.379572 & 2.6389 & 0.4202 & 0.047106 & -1.766066 & 3.5565 & 0.5662 \\
7 & 0.169371 & -3.729485 & 1.6830 & 0.2681 & 0.076445 & -2.698374 & 2.3276 & 0.3706 & 0.035825 & -2.085335 & 3.0126 & 0.4795 \\
8 & 0.167252 & -4.046426 & 1.5515 & 0.2471 & 0.065851 & -3.330929 & 1.8863 & 0.3002 & 0.023266 & -2.402870 & 2.6147 & 0.4162 \\
9 & 0.163398 & -4.524991 & 1.3877 & 0.2210 & 0.058556 & -3.804272 & 1.6514 & 0.2629 & 0.015007 & -2.561150 & 2.4532 & 0.3904 \\
\hline
\hline
\bottomrule
\end{tabular}
\label{tab:qnm_values}
\end{table*}

We first focus on the stability of the QNMs spectrum. Fig.~\ref{fig:qnm_evolution} presents the evolution of the complex QNM frequencies with the overtone index \(n\) under intrinsic coupled perturbations, with the core numerical results summarized in Table~\ref{tab:qnm_values}. The time evolution factor of the perturbation fields is \(e^{-i\omega t} = e^{\mathrm{Im}(\omega)t}e^{-i\mathrm{Re}(\omega)t}\). Hence, the imaginary part \(\mathrm{Im}(\omega)\) determines the stability of the perturbed black hole, while the oscillation behavior depends on the real part \(\mathrm{Re}(\omega)\). From Fig.~\ref{fig:qnm_evolution}(a), it is evident that for all overtone indices \(n \geq 0\), the eigenfrequencies satisfy \(\mathrm{Im}(\omega) < 0\). This means that the amplitude of the coupled metric-dilaton perturbations decays strictly exponentially over time, with no unstable modes that grow with time. This conclusion is qualitatively consistent with the results obtained under external scalar and spinor field perturbations~\cite{Gayen2023GRG}. It indicates that whether the MSW black hole is immersed in an external field or its intrinsic dynamical degrees of freedom are directly excited, the black hole solution is dynamically stable within the framework of the low-energy effective theory. This stability is a core support for the MSW black hole as a self-consistent solution in low-energy string theory. Furthermore, from Fig.~\ref{fig:qnm_evolution}(a), it can be seen that the fundamental mode \(n = 0\) has the smallest \(|\mathrm{Im}(\omega)|\), and the decay rate \(|\mathrm{Im}(\omega)|\) for higher overtones \(n \geq 1\) increases monotonically with the overtone index \(n\). This trend is entirely qualitatively consistent with the QNM evolution of four-dimensional Schwarzschild black hole and two-dimensional black hole under external field perturbations. Furthermore, pure two-dimensional Einstein gravity is topological and possesses no local propagating gravitational degrees of freedom; therefore, it cannot produce a non-trivial QNM spectrum. The dilaton field inherent in string theory introduces a local propagating dynamical degree of freedom into two-dimensional topological gravity. This degree of freedom is deeply coupled with the spacetime metric, together constituting the damped oscillation dynamics of the perturbed black hole. This is a key experimentally observable feature that distinguishes low-energy effective string gravity from pure two-dimensional Einstein gravity.

However, the dynamics of intrinsic perturbations exhibit qualitative differences from external field perturbations. As explored by Gayen and Koley, the QNM frequencies produced by external scalar field perturbations (even with non-minimal coupling) are purely imaginary (\(\operatorname{Re}(\omega) = 0\))~\cite{Gayen2023GRG}. This means that if an external observer uses a scalar field probe to perturb the MSW black hole, the observed black hole response is purely exponential decay, without any oscillatory component. For external spinor field perturbations, they did find QNM frequencies with non-zero real parts. Interestingly, these real parts are proportional to the coupling constant \(Q\) and are independent of the overtone index \(n\)~\cite{Gayen2023GRG}. Sebastian and Kuriakose, using the WKB method to study Dirac perturbations, also confirmed the existence of complex frequencies~\cite{Sebastian2014MPLA}. From Fig.~\ref{fig:qnm_evolution}(b), we can clearly observe that, qualitatively consistent with the spinor field case, all QNM eigenfrequencies for overtones \(n\geq 0\) of the intrinsically coupled perturbed 2D MSW black hole satisfy \(\operatorname{Re}(\omega) > 0\). This reveals that oscillatory behavior is not a universal response of the MSW spacetime to external probes, but rather originates from the specific coupling between the perturbing degrees of freedom and the spacetime geometry. In the external spinor field case, the oscillation arises from the interference effects of the Dirac field itself and its Yukawa-type coupling with the dilaton field. In contrast, in our intrinsic perturbation scenario, the oscillation directly stems from the coupling between the two dynamical degrees of freedom: the dilaton field and the metric field. Even in the complete absence of any external matter field, the spacetime geometry itself harbors the seeds of oscillation. The dilaton is no longer a passive probe detecting the black hole but is a dynamical partner deeply intertwined with the metric. Their coupling facilitates continuous energy exchange, thereby generating intrinsic oscillatory dynamics. In summary, the intrinsic perturbations of a two-dimensional stringy black hole are not a purely dissipative relaxation process but a dynamical process co-dominated by coupled oscillations and horizon dissipation.

Moreover, distinctly different from the spinor field results, we discover a non-monotonic behavior of the QNM real part with respect to \(n\), which precisely reflects the competition between two core effects in the intrinsically coupled system. As shown in Fig.~\ref{fig:qnm_evolution}(b), in the low-overtone region, the real part \(\operatorname{Re}(\omega)\) gradually increases with the overtone number. The spatial oscillation frequency of the perturbation rises, and the spatial overlap between metric fluctuations and dilaton fluctuations significantly increases. Their coupling strength is continuously amplified, and the energy exchange efficiency steadily improves. Consequently, the eigenfrequency \(\operatorname{Re}(\omega)\) rises rapidly with \(n\). Simultaneously, the decay rate \(|\operatorname{Im}(\omega)|\) for low overtones is relatively low, and the dissipative effect of the event horizon is insufficient to effectively suppress the trend of increasing coupled oscillation frequency. For high overtones, however, as \(|\operatorname{Im}(\omega)|\) increases linearly, the perturbation becomes highly localized near the black hole event horizon in the strong-gravity region. At this point, the horizon's dissipative effect begins to suppress the coupled oscillations, becoming the dominant dynamical factor and causing the effective oscillation frequency \(\operatorname{Re}(\omega)\) to decrease as \(n\) increases. This behavior starkly contrasts with the external spinor field spectrum, where the real part is completely independent of \(n\)~\cite{Gayen2023GRG}. This difference reflects a fundamental distinction in the coupling mechanisms. The oscillation frequency of the external spinor field is solely determined by its coupling constant \(Q\) with the dilaton and is independent of the mode's spatial structure. In contrast, the oscillation frequency of intrinsically coupled perturbations depends on the specific morphology of the mode wave function within the effective potential, resulting from the combined effects of coupling strength and spatial localization. This qualitative result reflects that the dilaton is not a passive background field but a dynamical degree of freedom deeply intertwined with the spacetime geometry. Its contribution to black hole perturbation dynamics is not only non-negligible but also varies significantly with the mode order. This phenomenon bears interesting similarities to the spectral behavior observed by Bl\'azquez-Salcedo et al. in higher-dimensional Einstein-Gauss-Bonnet-dilaton black hole~\cite{Blazquez2017PRD}, suggesting that dilaton coupling may lead to similar dynamical features in black hole across different dimensions.

On the other hand, we investigate the quantitative impact of the central charge parameter \(\sqrt{k}\) on the QNMs. As shown in Fig.~\ref{fig:qnm_evolution}(a), similar to the external field perturbation results, a larger central charge parameter \(\sqrt{k}\) leads to a smaller magnitude of the imaginary part \(|\operatorname{Im}(\omega)|\) of the QNM complex frequency. This is because, as seen in Fig.~\ref{fig:effpot}(b), a larger \(\sqrt{k}\) results in a lower potential barrier, making it easier for perturbation energy to leak out to infinity rather than being localized and dissipated near the horizon, thus prolonging the perturbation lifetime. Furthermore, from Fig.~\ref{fig:qnm_evolution}(b), we can discern that in the high-overtone region, due to the larger \(\sqrt{k}\) and consequently lower potential, the real part \(\operatorname{Re}(\omega)\) of the complex frequency decreases, meaning the oscillation frequency of the perturbation becomes smaller. It is worth noting that for external scalar field perturbations, the purely imaginary QNM frequency is directly proportional to a combination of \(\sqrt{k}\) and the mass term~\cite{Gayen2023GRG}. However, in intrinsic perturbations, this parameter dependence is more complexly encoded within the coupling matrix elements of the effective potential.

\begin{figure}
\centering
\subfigure[]
{\includegraphics[width=3in]{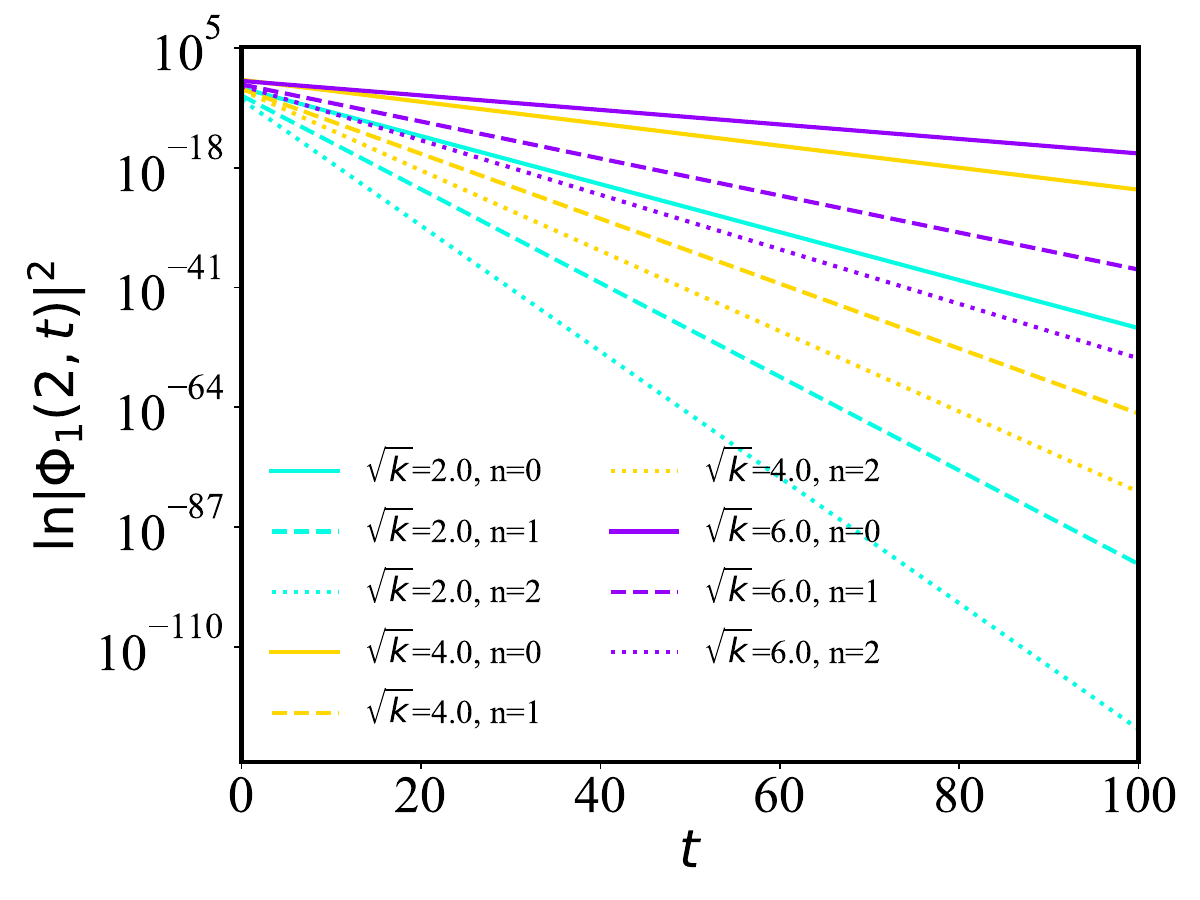}}
\subfigure[]
{\includegraphics[width=3in]{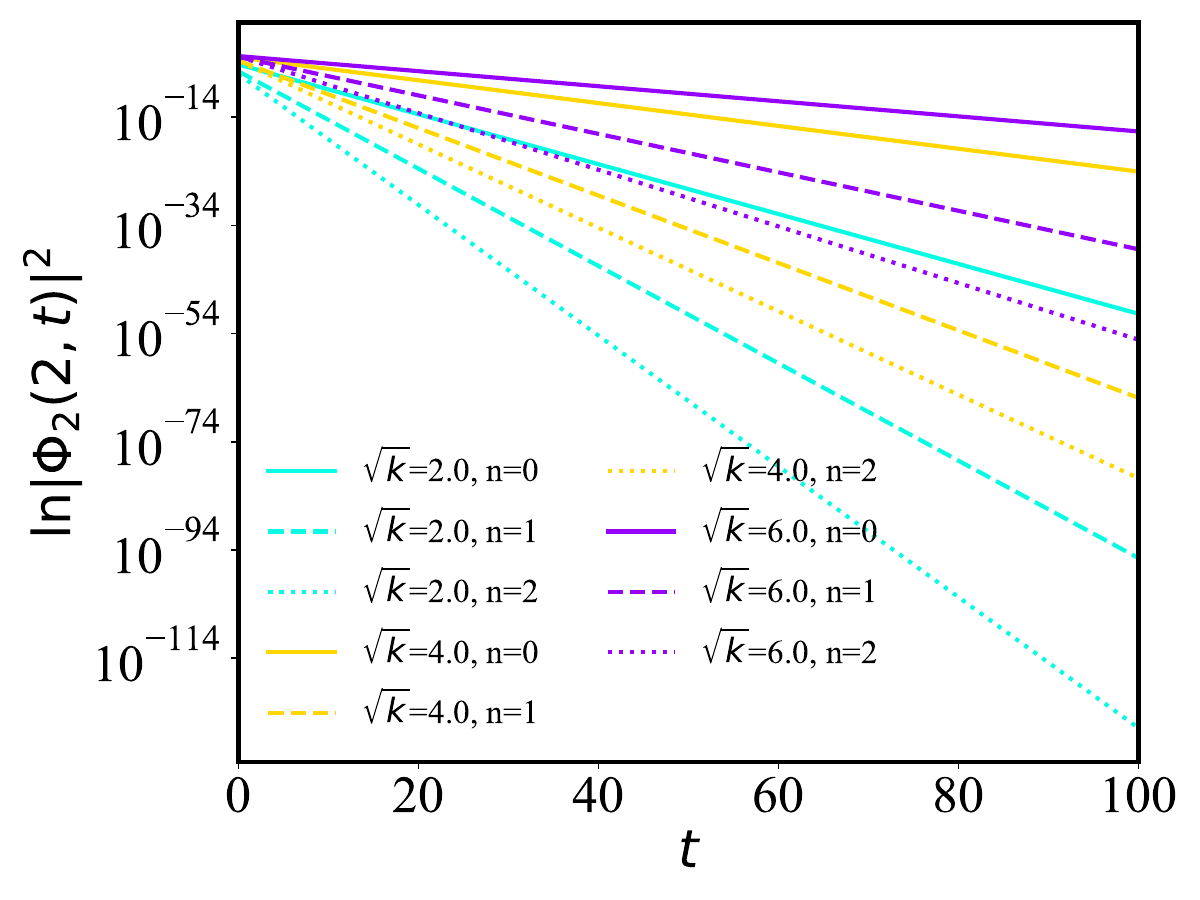}}
\caption{(Color online) Time evolution of the two branches of the perturbation wave function at \(r_* = 2.0\) for different parameters \(\sqrt{k}\) and modes \(n\).}
\label{fig:time_evolution}
\end{figure}

We can observe that although the real part of the QNM complex frequency is non-zero (\(\operatorname{Re}(\omega)\neq 0\)), it is very small, as shown in Table~\ref{tab:qnm_values}. Consequently, the damping ratio \(\eta = |\mathrm{Im}(\omega)| / \mathrm{Re}(\omega)\gg 1\). This leads to an oscillation period dominated by the real part that is much larger than the decay time \(T\gg \tau \equiv 1 / |\mathrm{Im}(\omega)|\), resulting in overdamped oscillations where the oscillatory behavior is effectively suppressed. To visually demonstrate the dynamical process of the black hole returning to a steady state after intrinsic perturbations, we plot in Fig.~\ref{fig:time_evolution} the time evolution of the squared magnitude of the two branches of the perturbation wave function, \(|\Phi_{1,2}(r_*,t)|^{2}\). In the calculation, we fix the tortoise coordinate at \(r_* = 2.0\) and select different central charge parameters \(\sqrt{k}\) and mode orders \(n\) to reveal the impact of parameter variations on the evolutionary behavior.

From Fig.~\ref{fig:time_evolution}, it can be clearly seen that in all cases, the squared modulus of the wave function decays monotonically with time, without any significant oscillatory structure. This feature is consistent with the fact, listed in Table~\ref{tab:qnm_values}, that the real part of the complex frequency is much smaller than the imaginary part (\(\operatorname{Re}(\omega)\ll |\operatorname{Im}(\omega)|\)). The system's damping ratio \(\eta = |\operatorname{Im}(\omega)| / \operatorname{Re}(\omega)\gg 1\), causing the oscillation period to be much larger than the decay time, resulting in overdamped evolution. In other words, although the coupling between the metric and dilaton introduces a non-zero real part (i.e., a tendency to oscillate), the dissipative effect of the event horizon is overwhelmingly dominant. This makes the overall response appear as a quasi-exponential decay, with the oscillatory component being completely submerged. Additionally, for a fixed mode order \(n\), the decay rate of the wave function slows down noticeably as \(\sqrt{k}\) increases. For example, the fundamental mode \(n = 0\) decays to near zero at about \(t\sim 5\) when \(\sqrt{k} = 2.0\), whereas it takes about \(t\sim 15\) when \(\sqrt{k} = 6.0\). This directly corresponds to the trend of decreasing \(|\operatorname{Im}(\omega)|\) with increasing \(\sqrt{k}\) shown in Table~\ref{tab:qnm_values}. The root cause is the lowering of the effective potential height with increasing \(\sqrt{k}\) as seen in Figure~\ref{fig:effpot}(b), which allows perturbations to leak more easily to infinity, reducing the energy localized near the horizon and thus slowing down the decay. From a string theory perspective, \(\sqrt{k}\) is related to the central charge of the 2D black hole. Smaller black hole (with larger \(\sqrt{k}/M\)) have flatter effective potentials, and their QNMs decay more slowly. This might suggest that the finiteness of the total number of microstates leads to observable modifications in macroscopic damping behavior. The influence of the mode order \(n\) is also significant: regardless of the value of \(\sqrt{k}\), the wave functions of higher overtones decay faster than those of lower overtones. Taking \(\sqrt{k} = 4.0\) as an example, the \(n = 0\) mode takes about \(t\sim 8\) to decay completely, while the \(n = 2\) mode only needs about \(t\sim 3\). This aligns with the monotonic increase of \(|\operatorname{Im}(\omega)|\) with \(n\) in Table~\ref{tab:qnm_values}, physically because higher modes are more localized near the horizon (where the effective potential peak is higher) and are thus more readily absorbed by the black hole. It is worth noting the subtle differences in the early-time evolution curves of the two branches \(\Phi_1\) and \(\Phi_2\), which subsequently synchronize. This reflects the coupling effect between metric and dilaton perturbations: initially, the energy distribution between the two differs, but they exchange energy through the coupling terms, eventually reaching a common decaying steady state. This brief ``pre-equilibrium" dynamics is a unique hallmark of intrinsically coupled systems, distinguishing them from external test fields, and further confirms that the dilaton field is not a passive background but a dynamical degree of freedom deeply bound to the spacetime geometry.

\begin{table}
\centering
\caption{The QNMs frequency spectrum and black hole behavior under external fields and intrinsic perturbations.}
\begin{tabular}{c|c c c} 
\hline\hline
Perturbation & $\mathrm{Im}(\omega)$ & $\mathrm{Re}(\omega)$ & Behavior  \\
\hline
Scalar & - & 0 & Decay \\
Spinor & - & + & Damped oscillation\\
Intrinsic & - & + & Overdamping\\
\hline\hline
\end{tabular}
\label{tab:comparison}
\end{table}

In summary, the comparative analysis with external field perturbations reveals a rich, multi-layered physical picture: the relaxation dynamics of the MSW black hole are highly sensitive to the nature of the perturbation. As shown in Table~\ref{tab:comparison}, the external scalar field probes purely dissipative decay modes; the external spinor field probes damped oscillatory modes, with the oscillation frequency solely determined by the coupling constant; the intrinsic coupled perturbations exhibit a non-zero real part, a non-monotonic evolution of the real part with \(n\), and overdamped behavior, profoundly reflecting the coupled nature of the dilaton and metric as a unified dynamical degree of freedom. These differences suggest that the quasinormal mode spectrum of a black hole is not a single, fixed "fingerprint," but rather acts like a multifaceted prism, revealing different facets of the black hole's microscopic dynamical structure from different angles (different perturbation channels). These results provide intuitive dynamical evidence for understanding the connection between the microscopic structure and macroscopic response of two-dimensional stringy black hole.

\section{Discussion and outlook}\label{secVI}

In this work, by solving the QNMs of intrinsic coupled perturbations of the metric and dilaton fields in the two-dimensional MSW black hole, we have obtained the characteristic features of its complex frequency spectrum. The results show that all modes satisfy \(\mathrm{Im}(\omega) < 0\), confirming the dynamical stability of the system. Simultaneously, the presence of a non-zero real part \(\mathrm{Re}(\omega) > 0\) signifies the excitation of coupled oscillations. The real part exhibits a non-monotonic variation with the overtone index \(n\)---increasing in the low-overtone region and decreasing in the high-overtone region---reflecting a competition mechanism between coupled oscillations and horizon dissipation. Nevertheless, intrinsic perturbations overall cause the black hole to exhibit overdamped evolutionary behavior. Based on this, we preliminarily discuss the following connections between black hole microstructure and macroscopic response.

The imaginary part of QNMs, \(|\mathrm{Im}(\omega)|\), representing the perturbation decay rate, possesses clear physical significance within a thermodynamic framework. For a black hole near equilibrium, its relaxation time \(\tau \equiv 1 / |\mathrm{Im}(\omega)|\) should be closely related to transport coefficients such as heat capacity and conductivity. In linear response theory, the decay rate is given by the poles of correlation functions, which in turn are connected to thermodynamic quantities via the fluctuation-dissipation theorem~\cite{Berti2009CQG,Kubo1966RPP}. Specifically, for a two-dimensional dilaton black hole, the heat capacity \(C\) can be obtained from the derivative of the ADM mass with respect to temperature, \(C = dM / dT_{H}\). Examining the fundamental mode decay rates for different \(\sqrt{k}\) in Table~\ref{tab:qnm_values}, we find that \(\tau\) increases with \(\sqrt{k}\). This behavior qualitatively aligns with the expectation from the harmonic oscillator statistical model, where an increase in the number of microscopic degrees of freedom leads to a more stable system~\cite{CADONI2002NPB}. If we assume that \(\sqrt{k}\) is positively correlated with the number of microscopic degrees of freedom \(N\) (in 2D CFT, the central charge \(c\) is proportional to \(N\)~\cite{Strominger1996PLB}), then the heat capacity \(C \propto N\) should also increase with \(\sqrt{k}\). However, establishing a quantitative relationship between \(\tau\) and heat capacity \(C\) requires further clarifying the connection between \(M\) and \(\sqrt{k}\) in thermodynamic processes. Since the definition of heat capacity involves the derivative of \(M\) with respect to \(T_{H}\), and \(T_{H}\) depends solely on \(\sqrt{k}\), without specifying how \(M\) varies with \(\sqrt{k}\), one cannot directly derive an explicit relation between \(C\) and \(\sqrt{k}\). This suggests that inferring thermodynamic properties from the QNM relaxation time necessitates considering a specific black hole thermodynamic ensemble, which will be a focus of subsequent work. Cadoni et al. have investigated the connection between quasinormal modes and microscopic descriptions of two-dimensional black hole, providing an important reference for this direction~\cite{CADONI2002NPB}.

Furthermore, the profound connection between the imaginary part of QNMs and black hole entropy can be understood through the concepts of adiabatic invariants and quantization conditions~\cite{Maggiore2008PRL}. For a one-dimensional Schr\"odinger-type equation, the Bohr-Sommerfeld quantization condition relates the action integral to energy levels. For dissipative systems, the appearance of complex frequencies implies that the action possesses an imaginary part, whose magnitude is inversely proportional to the lifetime of the state (i.e., the statistical fluctuation of entropy)~\cite{Shahar2007PRD}. Specifically, expanding the adiabatic invariant \(I = \oint \sqrt{\omega^{2} - V_{\mathrm{eff}}(r_{*})}\, dr_{*}\) near equilibrium, its real part yields energy quantization, while its imaginary part is related to sub-leading corrections to the entropy~\cite{Kunstatter2003PRL}. Recent work on the harmonic oscillator statistical model for two-dimensional dilaton black hole provides a concrete realization of this connection. Cadoni et al. proposed treating the black hole as a statistical ensemble composed of \(N\) decoupled harmonic oscillators with frequency \(\omega\). In the high-temperature limit, its Gibbs entropy recovers the Bekenstein-Hawking entropy \(S = 4\pi M / \sqrt{k}\), satisfying \(S = N\), meaning the number of microscopic degrees of freedom is directly determined by the entropy value~\cite{CADONI2002NPB}. This model also predicts a logarithmic correction term to the entropy, \(\Delta S \sim -\frac{1}{2} \ln (M T_{H})\), arising from finite-size effects~\cite{Carlip2000CQG}. From Table~\ref{tab:qnm_values}, it is evident that \(|\mathrm{Im}(\omega)|\) decreases as \(\sqrt{k}\) increases, and this non-linear decreasing trend may encode information about entropy corrections. If one can systematically couple the QNM imaginary part with background thermodynamic quantities such as temperature and heat capacity, it might be possible to establish a quantitative scaling relation to invert the number of microscopic degrees of freedom \(N\) from the frequency spectrum.

Inferring the number of microscopic degrees of freedom \(N\) from the QNM spectrum is feasible in principle, with the key lying in establishing a quantitative scaling relationship between \(|\mathrm{Im}(\omega)|\) and the central charge \(\sqrt{k}\). For a two-dimensional dilaton black hole, the Hawking temperature is \(T_{H} = (1/(4\pi))df/dr|_{r = r_{H}} = 1/(4\pi \sqrt{k})\), and the entropy is \(S = 4\pi M / \sqrt{k}\)~\cite{Gayen2023GRG,Mandal1991MPRA}. If we assume the fundamental mode decay rate obeys a scaling law \(|\mathrm{Im}(\omega_{0})| \sim T_{H} \cdot f(N)\), then by measuring \(|\mathrm{Im}(\omega_{0})|\) and \(T_{H}\), one could potentially infer \(N\). Our results preliminarily indicate that \(|\mathrm{Im}(\omega_{0})| / T_{H}\) slowly decreases as \(\sqrt{k}\) increases: for \(\sqrt{k} = 2.0\), \(|\mathrm{Im}(\omega_0)| / T_H \approx 0.531 / (1 / 8\pi) \approx 13.34\); for \(\sqrt{k} = 4.0\), this ratio decreases to \(0.243 / (1 / 16\pi) \approx 12.22\); and for \(\sqrt{k} = 6.0\), it further decreases to \(0.160 / (1 / 24\pi) \approx 12.07\). This trend suggests a possible scaling relation \(f(N) \sim N^{-\alpha}\), where \(\alpha\) is a positive constant. This observation is qualitatively consistent with the expectations of the harmonic oscillator statistical model~\cite{CADONI2002NPB}. This conjecture, however, requires verification through more precise numerical calculations and analytical derivations.

Although two-dimensional models are simplified in terms of dimensionality, their core physics—the coupling between the dilaton degree of freedom and geometry—is also present in higher-dimensional string theory black hole~\cite{HOROWITZ1991NPB}. If future gravitational wave observations could identify similar non-monotonic behavior or parameter-dependent features in QNM real parts, it might serve as a potential window to test low-energy effective string theory~\cite{Ferrari2020}. In summary, the results of this work provide preliminary dynamical evidence for understanding the connection between the microstructure and macroscopic response of string theory black hole. Through further theoretical development and deeper numerical investigations, it may be possible to extract more information about microscopic degrees of freedom from QNM spectra, paving the way towards ultimately unraveling the nature of black hole microstates.

\section{Summary}\label{secVII}

In this work, we have investigated the quasinormal modes associated with intrinsic coupled perturbations of the metric and dilaton fields in the two-dimensional MSW black hole. By introducing the field redefinition $\Phi=e^{-\phi}$ together with a conformal parametrization of the metric, the original gravity–dilaton action was reformulated in terms of the dynamical variables $\rho$ and $\Phi$. The resulting perturbation equations were reduced to a system of coupled Schr\"odinger-type equations, allowing the quasinormal-mode eigenvalue problem to be solved under the standard boundary conditions of purely ingoing waves at the horizon and purely outgoing waves at spatial infinity.

The computed spectrum shows that all modes satisfy $\mathrm{Im}(\omega)<0$, demonstrating the linear stability of the MSW black hole against intrinsic metric–dilaton perturbations. This result is consistent with previous studies of external scalar and spinor perturbations and further supports the dynamical robustness of this solution within the framework of low-energy effective string theory. A notable feature of the intrinsic perturbations is the appearance of nonvanishing real parts of the quasinormal frequencies. In contrast to external scalar perturbations, which produce purely imaginary spectra, the coupled metric–dilaton system exhibits oscillatory modes with $\mathrm{Re}(\omega)>0$. This behavior reflects the presence of genuine dynamical degrees of freedom arising from the dilaton field, which introduces a propagating scalar mode into otherwise topological two-dimensional gravity and enables energy exchange with the spacetime geometry.

Furthermore, the real part of the frequency displays a pronounced non-monotonic dependence on the overtone number. This feature can be interpreted as a manifestation of the competition between coupled oscillatory dynamics and horizon dissipation: low-overtone modes benefit from stronger metric–dilaton coupling, while higher overtones become increasingly dominated by near-horizon dissipative effects. The central-charge parameter $\sqrt{k}$ also plays an important role in shaping the spectrum. Increasing $\sqrt{k}$ systematically decreases the magnitude of the imaginary part of the frequencies, indicating slower damping and longer relaxation times. This behavior is consistent with the corresponding reduction of the effective potential barrier and suggests that the number of microscopic degrees of freedom influences the macroscopic relaxation dynamics of the black hole.

Overall, our results highlight that the quasinormal-mode spectrum of the MSW black hole is highly sensitive to the nature of the perturbation channel. Intrinsic metric–dilaton perturbations probe the internal dynamical structure of the spacetime and reveal characteristic features absent in external probe fields. These findings extend the understanding of quasinormal dynamics in low-dimensional stringy black hole and may provide a useful framework for exploring possible connections between black-hole relaxation processes and microscopic degrees of freedom in quantum gravity.

\begin{acknowledgments}
This work was supported by Natural Science Foundation of Jiangsu Province (grant no. BK20220122) and National Natural Science Foundation of China (grant no. 12233002).
\end{acknowledgments}

\end{CJK*}
\end{document}